%% file: main.tex
\documentclass[conference]{IEEEtran}
\IEEEoverridecommandlockouts
\usepackage{cite}
\usepackage{amsmath,amssymb,amsfonts}
\usepackage{algorithm-with-patch}
\usepackage{graphicx}
\usepackage{textcomp}
\usepackage{xcolor}
\usepackage{lipsum}

\usepackage{subcaption}
\usepackage[flushleft]{threeparttable}

\usepackage{booktabs}
\usepackage{bm}

\usepackage{tikz}
\input{zx.tikzstyles}
\input{zx.tikzdef}

\usetikzlibrary{arrows.meta}
\usetikzlibrary{calc,shapes.arrows}

\usepackage{tikzit}
\usepackage{quantikz}

\usepackage{MnSymbol}
\usepackage{orcidlink}

\usepackage[only,llbracket,rrbracket]{stmaryrd}
\SetSymbolFont{stmry}{bold}{U}{stmry}{m}{n} 
\newcommand{\intf}[1]{\left\llbracket{#1}\right\rrbracket}

\usepackage[capitalise]{cleveref}

\usepackage{microtype}

\usepackage[colorlinks=false]{hyperref}

\newcommand*\circled[1]{\tikz[baseline=(char.base)]{
            \node[shape=circle,draw,inner sep=0.5pt] (char) {#1};}}

\def\BibTeX{{\rm B\kern-.05em{\sc i\kern-.025em b}\kern-.08em
    T\kern-.1667em\lower.7ex\hbox{E}\kern-.125emX}}
\title{\textls[-8]{A Lazy Resynthesis Approach for Simultaneous T Gate and Two-Qubit Gate Optimization of Quantum Circuits}%
    \thanks{%
     \textcopyright 2025 IEEE.  Personal use of this material is permitted.  Permission from IEEE must be obtained for all other uses, in any current or future media, including reprinting/republishing this material for advertising or promotional purposes, creating new collective works, for resale or redistribution to servers or lists, or reuse of any copyrighted component of this work in other works.
     This work is supported by the National Science and Technology Council, R.O.C., Project Nos.: NSTC 113-2119-M-002-024 and NSTC 114-2119-M-002-020.
}
}
\author{%
\IEEEauthorblockN{%
Mu-Te Lau\IEEEauthorrefmark{1}\orcidlink{0009-0009-4729-7300}, Hsiang-Chun Yang, Hsin-Yu Chen, Chung-Yang (Ric) Huang\IEEEauthorrefmark{2}}%
\IEEEauthorblockA{%
Graduate Institute of Electronics Engineering, National Taiwan University \\
\IEEEauthorrefmark{1}mtlau@u.northwestern.edu, \IEEEauthorrefmark{2}cyhuang@ntu.edu.tw}%
}

\newcommand{\cx}{\mathrm{CX}}
\newcommand{\pauli}{\mathcal{P}}

\newcommand{\ctabl}{\mathcal{T}}
\newcommand{\ptabl}{\mathcal{S}}
\newcommand{\ztabl}{\mathcal{Z}}
\newcommand{\xtabl}{\mathcal{X}}

\begin{document}

\maketitle

\begin{abstract}
State-of-the-art quantum circuit optimization (QCO) algorithms for T-count reduction often lead to a substantial increase in two-qubit gate count (2Q-count)---a drawback that existing 2Q-count optimization techniques struggle to address effectively. In this work, we propose a novel lazy resynthesis approach for modern tableau-based QCO flows that significantly mitigates the 2Q-gate surges commonly introduced during T-count optimization in Clifford+T circuits. Experimental results show that our approach reduces 2Q-count overhead by 54.8\%, 15.3\%, and 68.0\% compared to tableau-based, ZX-calculus-based, and path-sum-based QCO algorithms, respectively. 
In terms of runtime, our method achieves speedups of 1.81$\times$ and 13.1$\times$ over the tableau-based and ZX-calculus-based methods, while performing comparably to the path-sum-based approach.
In summary, the proposed lazy resynthesis technique not only enhances the quality and performance of tableau-based QCO algorithms but also demonstrates superior efficiency and scalability compared to alternative QCO approaches such as ZX-calculus and path-sum-based techniques.
\end{abstract}


\input{src/1-intro}
\input{src/2-background}

\input{src/3-method}

\input{src/4-experiments}

\input{src/5-conclusion}

\section*{Acknowledgement}
This work is supported by the National Science and Technology Council, R.O.C., Project Nos.: NSTC 113-2119-M-002-024 and NSTC 114-2119-M-002-020.

\bibliographystyle{IEEEtran}
\bibliography{ref.bib}

\end{document}

%% file: zx.tikzstyles.tex
\tikzstyle{box}=[shape=rectangle, text height=1.5ex, text depth=0.25ex, yshift=0.5mm, fill=white, draw=black, minimum height=5mm, yshift=-0.5mm, minimum width=5mm, font={\small}]
\tikzstyle{Z dot}=[inner sep=0mm, minimum size=2mm, shape=circle, draw=black, fill={rgb,255: red,221; green,255; blue,221}]
\tikzstyle{Z phase dot}=[minimum size=5mm, font={\footnotesize\boldmath}, shape=rectangle, rounded corners=2mm, inner sep=0.2mm, outer sep=-2mm, scale=0.8, tikzit shape=circle, draw=black, fill={rgb,255: red,221; green,255; blue,221}, tikzit draw=blue]
\tikzstyle{X dot}=[Z dot, shape=circle, draw=black, fill={rgb,255: red,255; green,136; blue,136}]
\tikzstyle{X phase dot}=[Z phase dot, tikzit shape=circle, tikzit draw=blue, fill={rgb,255: red,255; green,136; blue,136}, font={\footnotesize\boldmath}]
\tikzstyle{hadamard}=[fill=yellow, draw=black, shape=rectangle, inner sep=0.6mm, minimum height=1.5mm, minimum width=1.5mm]
\tikzstyle{vertex}=[inner sep=0mm, minimum size=1mm, shape=circle, draw=black, fill=black]
\tikzstyle{vertex set}=[inner sep=0mm, minimum size=1mm, shape=circle, draw=black, fill=white, font={\footnotesize\boldmath}]

\tikzstyle{hadamard edge}=[-, dashed, dash pattern=on 2pt off 0.5pt, thick, draw={rgb,255: red,68; green,136; blue,255}]
\tikzstyle{brace edge}=[-, tikzit draw=blue, decorate, decoration={brace,amplitude=1mm,raise=-1mm}]
\tikzstyle{diredge}=[->]
\tikzstyle{dir hadamard edge}=[->, dashed, dash pattern=on 2pt off 0.5pt, thick, draw={rgb,255: red,68; green,136; blue,255}]

%% file: zx.tikzdef.tex
\usepackage{amsmath,amssymb,amsfonts}
\usepackage[colorlinks=false]{hyperref}
\usepackage{xcolor}
\usepackage{xspace}
\usepackage{bm}

\usetikzlibrary{decorations.pathmorphing}
\usetikzlibrary{fadings}
\usetikzlibrary{decorations.pathreplacing}
\usetikzlibrary{decorations.markings}


%% file: src/1-intro.tex
\section{Introduction} \label{sec:intro}
Minimizing the T-count (number of T gates) in Clifford+T quantum circuits is critical in fault-tolerant (FT) quantum computing, and state-of-the-art optimization algorithms have achieved substantial reductions in T-count. However, these reductions often come at the cost of significantly increased two-qubit gate counts (2Q-count), which can lead to more complex circuit structures \cite{vandaele2024lower,ruiz2024quantum,heyfron2018efficient}.

Traditionally, this increase in 2Q-count has been largely overlooked, as 2Q gates were assumed to be far less expensive than T gates in FT architectures \cite{ruiz2024quantum,heyfron2018efficient,de2020fast}, primarily due to the costly magic state distillation (MSD) required to produce resource states \cite{bravyi2005universal}. Recent developments, however, challenge this assumption. Improvements in MSD protocols and the emergence of alternative schemes \cite{litinski2019magic,gidney2019efficient,campbell2017unifying,bravyi2012magic,itogawa2024even} have considerably lowered the cost of T gates. In particular, a breakthrough reported in \cite{gidney2024magic} reduces the cost of T gates to the same order of magnitude as CNOT gates for small code distances (e.g.,  $d = 3, 5$) while maintaining a logical error rate of $6 \times 10^{-10}$, sufficient for practical quantum algorithms. Although more research is needed to determine if similar benefits extend to larger code distances, these findings indicate that the cost of 2Q gates is no longer negligible. Motivated by this shift, we aim to develop a Quantum Circuit Optimization (QCO) flow that mitigates the adverse impact of T-count optimization on 2Q-count.

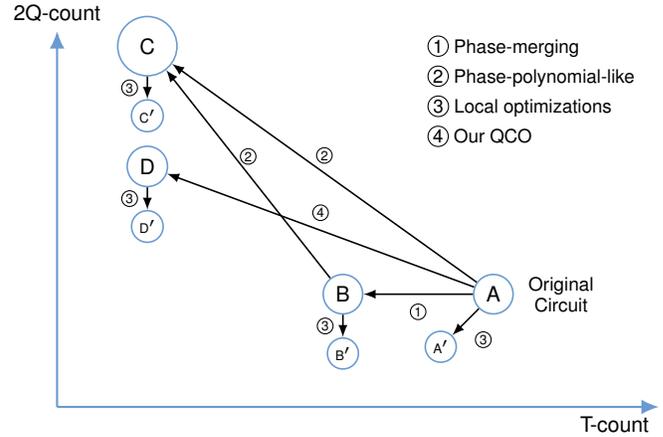
\begin{figure}
    \centering
    \input{figures/tikz/comparison-t-2q.tikz}
    \caption{Comparison on T-count and 2Q-count for various QCO techniques.}
    \label{fig:comparison-t-2q}
\end{figure}

To the best of our knowledge, we illustrate in \cref{fig:comparison-t-2q} a comprehensive comparison of trade-offs between T-count reduction and 2Q-count overhead across various modern QCO techniques. We categorize these methods into three classes based on how they handle T and 2Q gate optimizations:

\begin{itemize}
    \item \textbf{Phase-merging T-count optimizations} rearrange phase gates (not limited to T gates) into a sequence of Pauli rotations \cite{zhang2019optimizing} or phase gadgets in ZX-calculus \cite{kissinger2019reducing}. These gates are then possibly ``teleported'' and merged to form Clifford gates (e.g., $S$ or $I$) for elimination. Since these methods do not modify the placement or count of non-phase gates, the 2Q-count remains unchanged. This process corresponds to A $\to$ B in the figure.
    \item \textbf{Advanced phase-polynomial-like methods} offer more substantial T-count reductions (i.e., A $\to$ C) \cite{vandaele2024lower,ruiz2024quantum,heyfron2018efficient,de2020fast}. These methods convert the circuit into one or more phase polynomials (or equivalent representations) and reduce T-count by minimizing the number of monomial terms. However, they apply only to Hadamard-free (H-free) circuits, requiring preprocessing techniques like H-gadgetization or InternalHOpt \cite{vandaele2024optimal} to partition the original circuit into H-free subcircuits. As we will explain later, these T-count-optimized subcircuits are represented as intermediate forms such as Pauli rotation tableaux or ZX diagrams, necessitating a resynthesis step to convert them back into standard Clifford+T circuits \cite{vandaele2024optimal,kissinger2019reducing,amy2014polynomial,aaronson2004improved,markov2008optimal,maslov2018shorter,bravyi2021clifford,duncan2020graph,amy2017cnot,vandaele2022phase}. Due to the local nature of this resynthesis---applied independently to each H-free subcircuit---the reconstructed circuit may be far from globally optimal, often resulting in a significant 2Q-count increase.
    \item \textbf{Local optimization algorithms} apply rewrite rules to simplify circuits and reduce the number of gates, especially the redundant 2Q gates \cite{de2020fast,holker2023causal,xu2022quartz}. While these methods may incidentally reduce T-counts (e.g., A $\to$ A'), they typically do not yield additional T-count reductions beyond those achieved by specialized T-count optimizers. Their effects are illustrated as B $\to$ B', C $\to$ C', and D $\to$ D' in the figure.
\end{itemize}

In this paper, we propose a lazy resynthesis approach that builds upon several state-of-the-art quantum circuit optimization (QCO) algorithms to achieve optimal T-count reduction while minimizing the overhead of two-qubit (2Q) gates. Unlike traditional tableau-based QCO methods, which eagerly apply resynthesis to every Clifford and Pauli rotation tableau, our lazy strategy defers most resynthesis operations until the end of the flow. This postponement significantly reduces the 2Q gates introduced during intermediate resynthesis steps.

We implemented the proposed lazy resynthesis approach on Qsyn \cite{lau2024qsyn,qsynrepo}, our open-source C++ QCO platform that supports modern tableau-based, phase-polynomial-based, and ZX-calculus-based optimization techniques. We also benchmarked against the latest path-sum-based algorithms. Experimental results demonstrate that our approach reduces 2Q-count overhead by 54.8\%, 15.3\%, and 68.0\% compared to tableau-based, ZX-calculus-based, and path-sum-based QCO algorithms, respectively. Additionally, it achieves runtime improvements of 2.1$\times$, 4.4$\times$, and 2.59$\times$ over the above methods.

In summary, the proposed lazy resynthesis method significantly improves both the circuit quality and runtime performance of tableau-based QCO flows, while also exhibiting superior efficiency and scalability compared to alternative approaches such as ZX-calculus and path-sum-based techniques.

The rest of the paper is organized as follows. We first provide background on Pauli rotation and tableau-based QCO algorithms in \cref{sec:background} and then introduce our lazy resynthesis approach and the integrated QCO algorithm in \cref{sec:method}. The experimental results are presented in \cref{sec:experiments}. \cref{sec:conclusion} concludes the paper. 

%% file: figures/tikz/comparison-t-2q.tikz
\definecolor{myblue}{HTML}{6a9ad0}
\begin{tikzpicture}[
    >=latex,font=\sffamily\footnotesize,
    coordarrow/.style={draw=myblue,text=black,thick,->,-{Latex[width=2mm]}},
    plain/.style={black},
    main/.style={draw=myblue,text=black,circle,semithick,inner sep=0.25em},
    local/.style={draw=myblue,text=black,circle,semithick,inner sep=0.15em,font=\sffamily\tiny},
    opt/.style={->,semithick},
    optlabel/.style={midway,font=\sffamily\tiny},
    legend/.style={font=\sffamily\scriptsize,anchor=west}
]
    \draw[coordarrow] (0,0) -- (8,0) node[anchor=north east] {T-count};
    \draw[coordarrow] (0,0) -- (0,5) node[anchor=south] {2Q-count};

    \node[text width=1.35cm,align=center,font=\sffamily\scriptsize] at (6.7,1.5) {Original Circuit};
    \node[legend] at (4.8, 4.8) {\circled{1} Phase-merging};
    \node[legend] at (4.8, 4.4) {\circled{2} Phase-polynomial-like};
    \node[legend] at (4.8, 4.0) {\circled{3} Local optimizations};
    \node[legend] at (4.8, 3.6) {\circled{4} Our QCO};
    
    \node[main] (A) at (5.8,1.5) {A};
    \node[main] (B) at (3.8,1.5) {B};
    \node[main,inner sep=0.5em] (C) at (1.2,4.8) {C};
    \node[main] (D) at (1.2,3.2) {D};
    
    \node[local,below left=0.5 of A] (A') {A$'$};
    \node[local,below=0.3 of B] (B') {B$'$};
    \node[local,below=0.3 of C] (C') {C$'$};
    \node[local,below=0.3 of D] (D') {D$'$};

    \draw[opt] (A) -- (B) node[optlabel,below] {\circled{1}};
    \draw[->,semithick] (A) -- (C) node[optlabel,above] {\circled{2}};
    \draw[->,semithick] (B) -- (C) node[optlabel,above] {\circled{2}};
    \draw[->,semithick] (A) -- (D) node[optlabel,above] {\circled{4}};

    \draw[->,semithick] (A) -- (A') node[optlabel,below right] {\circled{3}};
    \draw[->,semithick] (B) -- (B') node[optlabel,left] {\circled{3}};
    \draw[->,semithick] (C) -- (C') node[optlabel,left] {\circled{3}};
    \draw[->,semithick] (D) -- (D') node[optlabel,left] {\circled{3}};
\end{tikzpicture}

%% file: src/2-background.tex
\section{Background and Related Work} \label{sec:background}

We assume the reader is familiar with basic quantum computing concepts, including quantum states, gates, circuits, Pauli operators/groups/strings, Clifford group, and Clifford+T gate set.

\subsection{Pauli Rotations}
Given a Pauli string $P$ in the $n$-qubit Pauli group $\mathcal{P}_n$, a Pauli rotation is defined as $R_P(\theta):= \exp (-i\theta P / 2)$. A Pauli rotation can be implemented using one phase gate $P(\theta)$ and some Clifford gates \cite{cowtan2019phase}. In a Clifford+T circuit, where the only non-Clifford is the T gate (equivalent to $R_Z(\pi/4)$), we fix $\theta = \pi / 4$ and simplify the notation to $R(P)$. Generally we use $R(P)$ to represent a Pauli rotation of Pauli string $P$ in a Clifford+T circuit.

Conjugating $R(P)$ with an arbitrary Clifford operator $C$, we have $CR(P)C^\dagger = R(CPC^\dagger)$, or equivelently, 
\begin{equation}
    CR(P) = R(CPC^\dagger)C. \label{eqn:pauli-rot-push}
\end{equation}
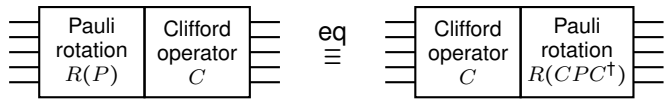
\begin{figure}[tb!]
    \centering
    \input{figures/tikz/pauli-rot-push.tikz}
    \caption{Circuit view of \cref{eqn:pauli-rot-push}}
    \label{fig:pauli-rot-push}
\end{figure}
In other words, as shown in \cref{fig:pauli-rot-push}, commuting the Clifford operator $C$ from the left side of the multiplication (i.e., $CR(P)$) to the right side (i.e., moving from the right side of $R(P)$ to the left side in the circuit) and replacing this $R(P)$ with a Pauli rotation of the conjugated Pauli string $CPC^\dagger$ preserves circuit equivalence.

Moreover, two Pauli rotations $R(P_1)$ and $R(P_2)$ commute if and only if (\textit{iff} below) the corresponding Pauli strings $P_1$ and $P_2$ commute:
\begin{equation*}
    [R(P_1), R(P_2)] = 0 \Leftrightarrow [P_1, P_2] = 0.
\end{equation*}

Lastly, since only $I$ and $Z$ are diagonal Pauli operators, we call $R(P)$ diagonal iff its Pauli string $P$ contains only $I$'s and $Z$'s. 

\subsection{Clifford and Pauli Rotation Tableaux}
Let $\pauli_n^* := \pm\{I,X,Y,Z\}^n \subseteq \pauli_n$ be a subset of the Pauli group $\pauli_n$. That is, $\pauli_n^*$ contains all the Pauli strings with leading coefficients $\pm 1$, not $\pm i$. We can encode such a Pauli string $P = (-1)^r P_1P_2...P_n \in \pauli_n^*$ as a $(2n+1) \times 1$ column bit-vector:
\begin{equation*}
    \begin{bmatrix}
        z_1 & \cdots & z_n & x_1 & \cdots & x_n & r
    \end{bmatrix}^\top,
\end{equation*}
where $z_i = 1$ iff $P_i = Z$ or $Y$, and $x_i = 1$ iff $P_i = X$ or $Y$.

Since any $n$-qubit Clifford operator can be uniquely represented with 2n Pauli strings in $\pauli_n^*$, we can utilize a $(2n+1) \times 2n$ binary matrix $\ctabl$, called Clifford or stabilizer tableau, to represent this Clifford operator \cite{aaronson2004improved,gottesman1998heisenberg}. Each column in this tableau corresponds to a Pauli string, and we can simulate Clifford operators (e.g., $\cx, S, H$) efficiently as certain row operations on $\ctabl$.

Similarly, $m$ consecutive Pauli rotations can be encoded into a $(2n+1) \times m$, almost-binary matrix $\ptabl$ called Pauli rotation tableau (PRT) \cite{vandaele2024optimal}, in which each column represents a Pauli rotation $R_P(\theta)$ of the form:
\begin{equation*}
    \begin{bmatrix}
        z_1 & \cdots & z_n & x_1 & \cdots & x_n & \theta
    \end{bmatrix}^\top.
\end{equation*}

Given a Clifford+T quantum circuit $\mathcal{C}$, since an adjacent Clifford operator and a phase gate (i.e., a Pauli Rotation) can be swapped in the quantum circuit based on \cref{eqn:pauli-rot-push}, we can rearrange all the Clifford and non-Clifford gates to the left and the right of the circuit, respectively. Letting $\ctabl$ and $\ptabl$ be the Clifford and Pauli rotation tableaux of the Clifford and non-Clifford circuit segments, we have $\intf{\mathcal{C}} = \intf\ptabl \intf\ctabl$, where $\intf{\cdot}$ denotes the unitary operator implemented by a quantum circuit or tableau.

\subsection{Tableau-based QCO}
In general, tableau-based T-count optimization algo-rithms consist of three steps: 
\begin{enumerate}
    \item Transforming the quantum circuit to Clifford and/or Pauli rotation tableaux
    \item Reducing T-count
    \item Extracting (resynthesizing) a quantum circuit from the optimized tableaux
\end{enumerate}
As discussed above, although we can construct Clifford and Pauli rotation tableaux via \cref{eqn:pauli-rot-push}, the tableau $\ptabl$ is often not diagonal, and its entries do not necessarily commute. In such cases, T-count reduction is limited to local phase merges (Point B in Figure 1).

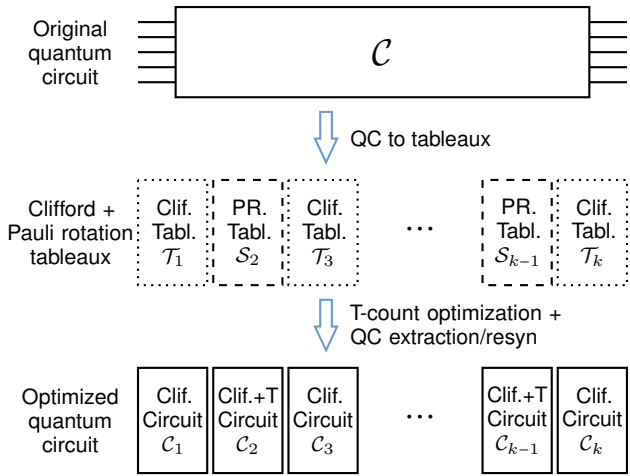
\begin{figure}[tb]
    \centering
    \input{figures/tikz/modern-t-count-opt-flow.tikz}
    \caption{Modern T-count optimization flow}
    \label{fig:modern-t-count-opt-flow}
\end{figure}
On the other hand, the recently proposed InternalHOpt algorithm \cite{vandaele2024optimal} addresses this by minimizing the Hadamard (H)-count, effectively partitioning the circuit into alternating Clifford and diagonal Pauli rotation tableaux (see \cref{fig:modern-t-count-opt-flow}). This enables the application of advanced T-count optimization techniques (e.g., phase polynomial-like \cite{ruiz2024quantum,vandaele2024lower,heyfron2018efficient,de2020fast}) on each diagonal tableau to achieve maximal T-count reduction (Point C in \cref{fig:comparison-t-2q}).

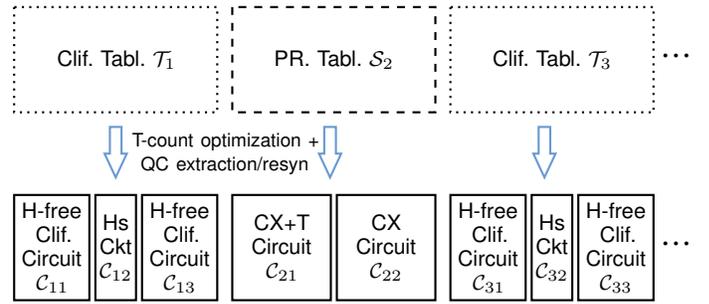
\begin{figure}[tb]
    \centering
    \input{figures/tikz/eager.tikz}
    \caption{Quantum circuit extraction in the T-count optimization flow (the eager resynthesis approach)}
    \label{fig:eager}
\end{figure}
The final step of the tableau-based T-count optimization flow is to extract the quantum circuit segments from the tableaux and cascade them as the final optimized circuit. For Clifford tableaux, we can apply a sequence of row operations that reduce the tableau to the identity and reconstruct the circuit in the reverse order \cite{kissinger2019reducing,vandaele2024optimal,aaronson2004improved,maslov2018shorter,bravyi2021clifford}. In particular, \cite{maslov2018shorter} shows that any Clifford circuit can be synthesized using one layer of Hadamard gates sandwiched between two H-free circuits (see \cref{fig:eager}).

For Pauli rotation tableaux, since they are now diagonal, the $X$ part in the tableau must be zero, and the remaining task is to manipulate the columns in $Z$ to be one-hot using CX gates. Once in the one-hot form, each column corresponds to a phase gate $P(\theta)$ that can be synthesized into the circuit and removed from the tableau. In this process, the parity terms accumulated on the qubits are resynthesized into a CX subcircuit (e.g., $C_{22}$ in \cref{fig:eager}) for uncomputing \cite{amy2014polynomial,amy2017cnot,vandaele2022phase}.

%% file: figures/tikz/pauli-rot-push.tikz
\begin{tikzpicture}[
    rectnode/.style={pos=.5,text width=1.35cm,align=center,font=\sffamily\footnotesize},
]
    \draw[thick] (-0.4, 0.2) -- (0, 0.2);
    \draw[thick] (-0.4, 0.4) -- (0, 0.4);
    \draw[thick] (-0.4, 0.6) -- (0, 0.6);
    \draw[thick] (-0.4, 0.8) -- (0, 0.8);
    \draw[thick] (-0.4, 1.0) -- (0, 1.0);
    \draw[thick] (0,0) rectangle ++(1.4,1.2) node[rectnode] {Pauli rotation $R(P)$};
    \draw[thick] (1.4,0) rectangle ++(1.4,1.2) node[rectnode] {Clifford operator $C$};
    \draw[thick] (2.8, 0.2) -- (3.2, 0.2);
    \draw[thick] (2.8, 0.4) -- (3.2, 0.4);
    \draw[thick] (2.8, 0.6) -- (3.2, 0.6);
    \draw[thick] (2.8, 0.8) -- (3.2, 0.8);
    \draw[thick] (2.8, 1.0) -- (3.2, 1.0);
    
    \node at (3.9, 0.7) {\Large\sffamily$\overset{\text{eq}}{\equiv}$};
    \draw[thick] (4.6, 0.2) -- (5, 0.2);
    \draw[thick] (4.6, 0.4) -- (5, 0.4);
    \draw[thick] (4.6, 0.6) -- (5, 0.6);
    \draw[thick] (4.6, 0.8) -- (5, 0.8);
    \draw[thick] (4.6, 1.0) -- (5, 1.0);
    \draw[thick] (5,0) rectangle ++(1.4,1.2) node[rectnode] {Clifford operator $C$};
    \draw[thick] (6.4,0) rectangle ++(1.5,1.2) node[rectnode] {Pauli rotation $R(CPC^\dagger)$};
    \draw[thick] (7.9, 0.2) -- (8.3, 0.2);
    \draw[thick] (7.9, 0.4) -- (8.3, 0.4);
    \draw[thick] (7.9, 0.6) -- (8.3, 0.6);
    \draw[thick] (7.9, 0.8) -- (8.3, 0.8);
    \draw[thick] (7.9, 1.0) -- (8.3, 1.0);
\end{tikzpicture}

%% file: figures/tikz/modern-t-count-opt-flow.tikz
\definecolor{myblue}{HTML}{6a9ad0}
\begin{tikzpicture}[
    yscale=-1,font=\sffamily\footnotesize,
    capt/.style={align=center,font=\sffamily\footnotesize},
    rectnode/.style={pos=.5,text width=1.35cm,capt},
    fat arrow/.style={every to/.style={
      to path={
        let \p1 = ($(\tikztotarget)-(\tikztostart)$),
            \n1 = {veclen(\x1,\y1)},
            \n2 = {mod(scalar(atan2(\y1,\x1))+360, 360)} 
        in
        -- (\tikztotarget)
        node[draw, myblue,
             inner xsep=0pt,inner ysep=1.8pt, 
             minimum height=\n1-\pgflinewidth,
             single arrow,
             rotate=-\n2, 
             anchor=tip, 
             single arrow head extend=2pt,
             single arrow tip angle=50,
             #1          
             ]
          {} \tikztonodes}
    }},
]
    \node[capt,text width=1.7cm] at (-1.4, 0.6) {Original quantum circuit};
    \node[capt,text width=1.7cm] at (-1.4, 3) {Clifford + Pauli rotation tableaux};
    \node[capt,text width=1.7cm] at (-1.4, 5.5) {Optimized quantum circuit};
    
    \draw[thick] (-0.5, 0.2) -- (0, 0.2);
    \draw[thick] (-0.5, 0.4) -- (0, 0.4);
    \draw[thick] (-0.5, 0.6) -- (0, 0.6);
    \draw[thick] (-0.5, 0.8) -- (0, 0.8);
    \draw[thick] (-0.5, 1.0) -- (0, 1.0);
    \draw[thick] (0,0) rectangle ++(5.5,1.2) node[rectnode] {\Large$\mathcal{C}$};
    \draw[thick] (5.5, 0.2) -- (6, 0.2);
    \draw[thick] (5.5, 0.4) -- (6, 0.4);
    \draw[thick] (5.5, 0.6) -- (6, 0.6);
    \draw[thick] (5.5, 0.8) -- (6, 0.8);
    \draw[thick] (5.5, 1.0) -- (6, 1.0);

    \path[fat arrow, thick] (2,1.4) to (2,2.1);
    \node[anchor=west] at (2.2,1.75) {QC to tableaux};
    
    \draw[thick,dotted] (-0.5,2.3) rectangle ++(0.92,1.4) node[rectnode,text width=0.8cm] {Clif. Tabl. $\ctabl_1$};
    \draw[thick,dashed] (0.5,2.3) rectangle ++(0.92,1.4) node[rectnode,text width=0.8cm] {PR. Tabl. $\ptabl_2$};
    \draw[thick,dotted] (1.5,2.3) rectangle ++(0.92,1.4) node[rectnode,text width=0.8cm] {Clif. Tabl. $\ctabl_3$};
    \node at (3.25, 3) {\Large$\cdots$};
    \draw[thick,dashed] (4.08,2.3) rectangle ++(0.92,1.4) node[rectnode,text width=0.8cm] {PR. Tabl. $\ptabl_{k-1}$};
    \draw[thick,dotted] (5.08,2.3) rectangle ++(0.92,1.4) node[rectnode,text width=0.8cm] {Clif. Tabl. $\ctabl_k$};

    \path[fat arrow, thick] (2,3.9) to (2,4.6);
    \node[anchor=west,text width=2.8cm] at (2.2,4.25) {T-count optimization + QC extraction/resyn};
    
    \draw[thick] (-0.5,4.8) rectangle ++(0.92,1.4) node[rectnode,text width=0.8cm] {Clif. Circuit $\mathcal{C}_1$};
    \draw[thick] (0.5,4.8) rectangle ++(0.92,1.4) node[rectnode,text width=0.8cm] {Clif.+T Circuit $\mathcal{C}_2$};
    \draw[thick] (1.5,4.8) rectangle ++(0.92,1.4) node[rectnode,text width=0.8cm] {Clif. Circuit $\mathcal{C}_3$};
    \node at (3.25, 5.5) {\Large$\cdots$};
    \draw[thick] (4.08,4.8) rectangle ++(0.92,1.4) node[rectnode,text width=0.8cm] {Clif.+T Circuit $\mathcal{C}_{k-1}$};
    \draw[thick] (5.08,4.8) rectangle ++(0.92,1.4) node[rectnode,text width=0.8cm] {Clif. Circuit $\mathcal{C}_k$};
\end{tikzpicture}

%% file: figures/tikz/eager.tikz
\definecolor{myblue}{HTML}{6a9ad0}
\begin{tikzpicture}[
    yscale=-1,font=\sffamily\footnotesize,
    capt/.style={align=center,font=\sffamily\footnotesize},
    rectnode/.style={pos=.5,text width=1.35cm,capt},
    fat arrow/.style={every to/.style={
      to path={
        let \p1 = ($(\tikztotarget)-(\tikztostart)$),
            \n1 = {veclen(\x1,\y1)},
            \n2 = {mod(scalar(atan2(\y1,\x1))+360, 360)} 
        in
        -- (\tikztotarget)
        node[draw, myblue,
             inner xsep=0pt,inner ysep=1.8pt, 
             minimum height=\n1-\pgflinewidth,
             single arrow,
             rotate=-\n2, 
             anchor=tip, 
             single arrow head extend=2pt,
             single arrow tip angle=50,
             #1          
             ]
          {} \tikztonodes}
    }},
]
    \draw[thick,dotted] (0,0) rectangle ++(2.7,1.4) node[rectnode,text width=2cm] {Clif. Tabl. $\ctabl_1$};
    \draw[thick,dashed] (2.9,0) rectangle ++(2.7,1.4) node[rectnode,text width=2cm] {PR. Tabl. $\ptabl_2$};
    \draw[thick,dotted] (5.8,0) rectangle ++(2.7,1.4) node[rectnode,text width=2cm] {Clif. Tabl. $\ctabl_3$};
    \node at (8.8,0.7) {\Large$\cdots$};

    \path[fat arrow, thick] (1.35,1.6) to (1.35,2.3);
    \node[align=center, text width=4.7cm] at (2.8,1.95) {\scriptsize T-count optimization +\\QC extraction/resyn};
    \path[fat arrow, thick] (4.2,1.6) to (4.2,2.3);
    \path[fat arrow, thick] (7.05,1.6) to (7.05,2.3);

    \draw[thick] (0,2.5) rectangle ++(1,1.4) node[rectnode,text width=0.8cm] {H-free Clif. Circuit $\mathcal{C}_{11}$};
    \draw[thick] (1.08,2.5) rectangle ++(0.54,1.4) node[rectnode,text width=0.8cm] {Hs Ckt $\mathcal{C}_{12}$};
    \draw[thick] (1.7,2.5) rectangle ++(1,1.4) node[rectnode,text width=0.8cm] {H-free Clif. Circuit $\mathcal{C}_{13}$};

    \draw[thick] (2.9,2.5) rectangle ++(1.31,1.4) node[rectnode,text width=0.8cm] {CX+T Circuit $\mathcal{C}_{21}$};
    \draw[thick] (4.29,2.5) rectangle ++(1.31,1.4) node[rectnode,text width=0.8cm] {CX Circuit $\mathcal{C}_{22}$};

    \draw[thick] (5.8,2.5) rectangle ++(1,1.4) node[rectnode,text width=0.8cm] {H-free Clif. Circuit $\mathcal{C}_{31}$};
    \draw[thick] (6.88,2.5) rectangle ++(0.54,1.4) node[rectnode,text width=0.8cm] {Hs Ckt $\mathcal{C}_{32}$};
    \draw[thick] (7.5,2.5) rectangle ++(1,1.4) node[rectnode,text width=0.8cm] {H-free Clif. Circuit $\mathcal{C}_{33}$};
    \node at (8.8,3.2) {\Large$\cdots$};
\end{tikzpicture}

%% file: src/3-method.tex
\section{Lazy Resynthesis Approach in Tableau-Based Quantum Circuit Optimization} \label{sec:method}
As discussed in the previous section, modern tableau-based quantum circuit optimization (QCO) flows perform circuit extraction eagerly on each tableau, applying full resynthesis before moving to the next. We refer to this as the eager resynthesis approach, in which resynthesis occurs as early as possible. As we will explain, this eager strategy is a major contributor to the increase in 2Q-count observed in tableau-based QCO algorithms.

In contrast, we propose a lazy resynthesis approach, which defers portions of Clifford circuit resynthesis by strategically merging Clifford operators with subsequent tableaux. As our experimental results show, this deferred strategy effectively reduces the overhead of 2Q gates.

\subsection{The Reason for the Surge of 2Q-Count}

In tableau-based QCO, quantum circuits are extracted from Clifford and Pauli rotation tableaux by performing row operations that simplify parts of the tableau into identities, one-hot forms, or zero matrices. For convenience, we refer to the submatrices formed by a tableau's $z$ entries and $x$ entries as $\ztabl$ and $\xtabl$, respectively. Accordingly, $\ztabl_i$ and $\xtabl_i$ denote the $i$th rows of these submatrices. 

The following are common row operations used during circuit extraction:
\begin{itemize}
    \item Swapping $\ztabl_i$ with the corresponding $\xtabl_i$
    \item Adding $\xtabl_i$ to $\ztabl_i$
    \item Simultaneously adding $\ztabl_t$ to $\ztabl_c$, and $\xtabl_c$ to $\xtabl_t$
\end{itemize}

Up to some Pauli differences that are easy to correct, these operations correspond to synthesizing H, S, and CX gates, respectively \cite{aaronson2004improved}.

As illustrated in \cref{fig:eager}, subcircuits such as $\mathcal{C}_{11}, \mathcal{C}_{13}, \mathcal{C}_{21}, \mathcal{C}_{22}, \mathcal{C}_{31}, \mathcal{C}_{33}$... etc., contain numerous CX gates introduced during local tableau resynthesis. Because each tableau is processed in isolation---without considering adjacent tableaux---many of these CX gates can be optimized away in the global context. Although local optimization techniques \cite{de2020fast,holker2023causal,xu2022quartz} can mitigate some of this overhead, the reduction is limited, and the total 2Q-count often increases significantly compared to the original unoptimized circuit.
\begin{figure}
    \centering
    \input{figures/tikz/lazy.tikz}
    \caption{An illustrative example of the proposed lazy resynthesis approach}
    \label{fig:lazy}
\end{figure}
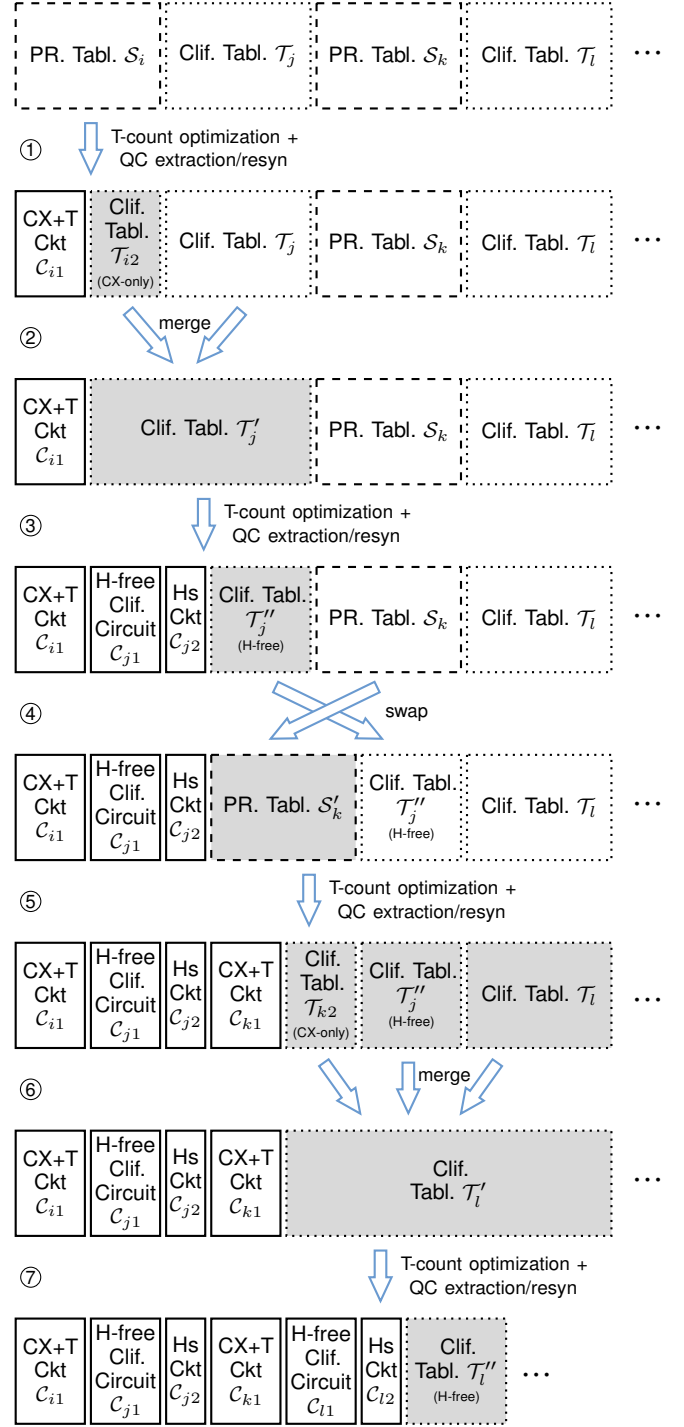
\subsection{The Proposed Lazy-Resynthesis Approach}
To clearly explain our method, we present the lazy resynthesis flow step by step, as illustrated in \cref{fig:lazy}. In this diagram:
\begin{itemize}
    \item Pauli rotation and Clifford tableaux are depicted as dashed and dotted blocks, respectively;
    \item Extracted sub-circuits are shown as solid blocks.
\end{itemize}

Without loss of generality, as we perform the circuit extraction alternately on Pauli rotation and Clifford tableaux, conventionally from left (i.e. circuit input) to right (i.e. circuit output), let's assume the QCO algorithm is optimizing on the Pauli rotation tableau $\ptabl_i$, followed by tableaux $\ctabl_j$, $\ptabl_k$ and $\ctabl_l$.

In Steps \circled{1} and \circled{2}, when we perform the phase-polynomial-like QCO algorithm on $\ptabl_i$, we construct the CX+T circuit only (i.e., $\mathcal{C}_{i1}$) and not the uncomputing circuit. Instead, we apply the CX-only conjugations as well as Pauli byproducts as row operations on the next Clifford tableau $\ctabl_j$. This effectively merges the uncomputing Clifford tableau into $\ctabl_j$, producing an updated Clifford tableau $\ctabl_j'$.

We then perform Clifford circuit optimization on $\ctabl_j'$ (Step \circled{3}). According to \cite{maslov2018shorter} and \cite{vandaele2024optimal}, we can first create an H-free sub-circuit $\mathcal{C}_{j1}$ and a layer of H gates $\mathcal{C}_{j2}$ by applying row operations on $\ctabl_j'$. This will result in a new Clifford tableau $\ctabl_j''$, which also corresponds to an H-free subcircuit.

However, different from the previous QCO algorithms where the Clifford circuit for $\ctabl_j''$ is immediately extracted on the spot, we instead delay the resynthesis of the Clifford tableau  and swap it with the adjacent Pauli rotation tableau $\ptabl_k$. According to \cref{eqn:pauli-rot-push}, $\ptabl_k$ will be updated to $\ptabl_k'$ by Clifford conjugations (Step \circled{4}).

In Step \circled{5}, we continue applying the T-count optimization on the Pauli rotation tableau $\ptabl_k'$ and extract the quantum circuit as in Step \circled{1}. However, we then merge the deferred Clifford tableau $\ctabl_{k2}$ with the swapped Clifford tableau $\ctabl_j''$ in Step \circled{4} and the next optimization target $\ctabl_l$ (Step \circled{6}).

These steps will be repeated on the remaining tableaux until the entire Clifford+T circuit is optimized.

\subsection{2Q-Count and Time Complexity Analysis} \label{subsect:complexity}

Comparing the previous eager and our proposed lazy resynthesis approaches, as shown in \cref{fig:eager,fig:lazy}, respectively, we can see that the resynthesis of the H-free Clifford subcircuits (e.g., $\mathcal{C}_{13}$ and $\mathcal{C}_{33}$ in \cref{fig:eager}) is deferred, and their Clifford tableaux (e.g., $\ctabl_j''$ in \cref{fig:lazy}) are swapped with the adjacent Pauli rotation tableaux (e.g., $\ptabl_k$ in \cref{fig:lazy}) and later merged with the next Clifford tableaux (e.g., $\ctabl_l$ in \cref{fig:lazy}). Likewise, the resynthesis of the CX-only uncomputing subcircuits (e.g., $\mathcal{C}_{22}$ in \cref{fig:eager}) is also deferred, and their Clifford tableaux (e.g., $\ctabl_{i2}$ in \cref{fig:lazy}) are merged with the adjacent Clifford tableaux (e.g., $\ctabl_j$ in \cref{fig:lazy}).

Let $n$ be the number of qubits in the quantum circuit, $t$ be the T-count or the number of Pauli rotations in the tableaux, and $h$ be the number of Hadamard gates. We can estimate the number of synthesized 2Q gates in the circuit extraction process as follows. 

\subsubsection{\textbf{2Q-Count Complexity Analysis}}

For the eager resynthesis approach, since the number of Hadamard gates ($h_{i1}$) generated in diagonalizing each Clifford tableau ($\ctabl_i$) is equal to $\mathrm{rank}(\ctabl_i)$ \cite{vandaele2024optimal} and each diagonalization takes $O(n)$ row operations (as CXs), we have:
\begin{equation}
\begin{split}
    &\text{2Q-count for the diagonalization subcircuits} \\
    &\text{(e.g., $\mathcal{C}_{11}$ and $\mathcal{C}_{31}$ in \cref{fig:eager})} = O(nh).
\end{split}
\label{eqn:2q-eager-1}
\end{equation}

From \cite{aaronson2004improved,markov2008optimal}, we know that it takes $O(n^2/\log n)$ CX gates to resynthesize a linear Clifford circuit. In addition, as there is at least one Hadamard gate in the one-layer H-circuit for each of the Clifford tableaux, the number of the resynthesized H-free subcircuits (e.g., $\mathcal{C}_{13}$ and $\mathcal{C}_{31}$ in \cref{fig:eager}) must be less than $h$. Therefore, we have:
\begin{equation}
\begin{split}
    &\text{2Q-count for the H-free subcircuits} \\
    &\text{(e.g., $\mathcal{C}_{13}$ and $\mathcal{C}_{33}$ in \cref{fig:eager})} = O(n^2h/\log n).
\end{split}
\label{eqn:2q-eager-2}
\end{equation}

For the Pauli rotation tableaux, since it takes $O(n)$ CX gates to synthesize each monomial phase term (corresponding to a phase gate) \cite{meijer2024advances}, we have:
\begin{equation}
\begin{split}
    &\text{2Q-count for the Pauli rotation subcircuits} \\
    &\text{(e.g., $\mathcal{C}_{21}$ in \cref{fig:eager})} = O(nt).
\end{split}
\label{eqn:2q-eager-3}
\end{equation}

Lastly, to uncompute the Clifford operators of the synthesized Pauli Pauli rotation subcircuits, we need $O(n^2/\log n)$ CX gates for each Pauli rotation subcircuits. Again noting that there are at most $O(h)$ such subcircuits, we have:
\begin{equation}
\begin{split}
    &\text{2Q-count for the uncomputing subcircuits} \\
    &\text{(e.g., $\mathcal{C}_{22}$ in \cref{fig:eager})} = O(n^2h/\log n).
\end{split}
\label{eqn:2q-eager-4}
\end{equation}

From \cref{eqn:2q-eager-1,eqn:2q-eager-2,eqn:2q-eager-3,eqn:2q-eager-4}, we can conclude:
\begin{equation}
\begin{split}
    &\text{2Q-count for the eager resynthesis approach} \\
    &= O(n^2h/\log n + nt + nh).
\end{split}
\label{eqn:2q-eager-5}
\end{equation}

On the other hand, for our proposed lazy resynthesis approach, since we do not perform the resynthesis from some of the Clifford tableaux (e.g., $\mathcal{C}_{13},\mathcal{C}_{33}$ and $\mathcal{C}_{22}$ in \cref{fig:eager}) and defer it till the last Clifford tableau, our 2Q-count can be computed as
\begin{equation}
\begin{split}
    &\text{2Q-count for the lazy resynthesis approach} \\
    &= O(n^2/\log n + nt + nh).
\end{split}
\label{eqn:2q-lazy}
\end{equation}

Comparing \cref{eqn:2q-eager-5,eqn:2q-lazy}, our proposed lazy resynthesis approach can improve the 2Q-count overhead as many as $O(h)$ times, asymptotically.

\subsubsection{\textbf{Time Complexity Analysis}}
We assume that the runtime is proportional to the addition of (a) the number of row operations for synthesizing 2Q gates, and (b) the counts to merge and swap the tableaux. Since it takes $O(n)$ and $O(t)$ to perform each row operation on Clifford and Pauli rotation tableaux, we have
\begin{equation}
\begin{split}
    &\text{2Q-count for the eager resynthesis approach} \\
    &= O(n^3h/\log n + nt^2 + n^2h).
\end{split}
\label{eqn:time-eager}
\end{equation}

On the other hand, for our proposed lazy resynthesis approach, resynthesizing the tableaux takes:
\begin{equation}
\begin{split}
    &\text{2Q-count for lazily resynthesizing 2Q gates} \\
    &= O(n^3/\log n + nt^2 + n^2h).
\end{split}
\label{eqn:time-lazy-resyn}
\end{equation}

In addition, to merge two Clifford tableaux, the runtime is proportional to number of row operations to diagonalize the matrices, which is of the same time complexity for resynthesis from a tableau (i.e., $O(n^3/\log n)$). Since there are at most $O(h)$ times of Clifford tableau merges, we have that
\begin{equation}
\begin{split}
    &\text{2Q-count for merging Clifford tableaux} \\
    &= O(n^3h/\log n).
\end{split}
\label{eqn:time-lazy-merge}
\end{equation}

Finally, we calculate the time complexity for swapping the Clifford and Pauli rotation tableaux. Since swapping a Clifford tableau with a Pauli rotation tableau works by conjugating the Pauli rotation tableau with the Clifford operator, we will first extract a string of Clifford operators from the Clifford tableau, which again takes $O(n^3/\log n)$ for at most $O(h)$ Clifford tableaux, before applying them to the Pauli rotation tableaux, which takes $O(nt^2/\log n)$. Therefore,
\begin{equation}
\begin{split}
    &\text{2Q-count for swapping Clifford and Pauli rotation} \\
    &\text{tableaux} = O((n^3h + nt^2)/\log n).
\end{split}
\label{eqn:time-lazy-swap}
\end{equation}

Putting \cref{eqn:time-lazy-resyn,eqn:time-lazy-merge,eqn:time-lazy-swap} together, we get
\begin{equation}
\begin{split}
    &\text{2Q-count for the lazy resynthesis approach} \\
    &= O(n^3h/\log n + nt^2 + n^2h).
\end{split}
\label{eqn:time-lazy}
\end{equation}

In summary, our algorithm synthesizes asymptotically fewer 2Q gates while maintaining the same time complexity as prior methods. This improvement is a direct result of the lazy resynthesis strategy. It is worth noting that these theoretical bounds for both 2Q-count and time complexity and time complexity are relatively loose; in practice, the lazy approach consistently delivers not only 2Q-counts but also better runtime performance.

%% file: figures/tikz/lazy.tikz
\definecolor{myblue}{HTML}{6a9ad0}
\begin{tikzpicture}[
    yscale=-1,font=\sffamily\footnotesize,
    capt/.style={align=center,font=\sffamily\footnotesize},
    rectnode/.style={pos=.5,text width=1.35cm,capt},
    fat arrow/.style={every to/.style={
      to path={
        let \p1 = ($(\tikztotarget)-(\tikztostart)$),
            \n1 = {veclen(\x1,\y1)},
            \n2 = {mod(scalar(atan2(\y1,\x1))+360, 360)} 
        in
        -- (\tikztotarget)
        node[draw, myblue, fill=white,
             inner xsep=0pt,inner ysep=1.8pt, 
             minimum height=\n1-\pgflinewidth,
             single arrow,
             rotate=-\n2, 
             anchor=tip, 
             single arrow head extend=2pt,
             single arrow tip angle=50,
             #1          
             ]
          {} \tikztonodes}
    }},
]
    \draw[thick,dashed] (0,0) rectangle ++(1.92,1.4) node[rectnode,text width=2cm] {PR. Tabl. $\ptabl_i$};
    \draw[thick,dotted] (2,0) rectangle ++(1.92,1.4) node[rectnode,text width=2cm] {Clif. Tabl. $\ctabl_j$};
    \draw[thick,dashed] (4,0) rectangle ++(1.92,1.4) node[rectnode,text width=2cm] {PR. Tabl. $\ptabl_k$};
    \draw[thick,dotted] (6,0) rectangle ++(1.92,1.4) node[rectnode,text width=2cm] {Clif. Tabl. $\ctabl_l$};
    \node at (8.4,0.7) {\Large$\cdots$};

    \node at (0.2,1.95) {\scriptsize\circled{1}};
    \path[fat arrow, thick] (1,1.6) to (1,2.3);
    \node[align=center, text width=4.7cm] at (2.5,1.95) {\scriptsize T-count optimization +\\QC extraction/resyn};

    \draw[thick] (0,2.5) rectangle ++(0.92,1.4) node[rectnode,text width=0.85cm] {CX+T Ckt $\mathcal{C}_{i1}$};
    \draw[thick,dotted,fill=gray!30] (1,2.5) rectangle ++(0.92,1.4) node[rectnode,text width=0.85cm] {Clif. Tabl. $\ctabl_{i2}$ {\tiny (CX-only)}};
    \draw[thick,dotted] (2,2.5) rectangle ++(1.92,1.4) node[rectnode,text width=2cm] {Clif. Tabl. $\ctabl_j$};
    \draw[thick,dashed] (4,2.5) rectangle ++(1.92,1.4) node[rectnode,text width=2cm] {PR. Tabl. $\ptabl_k$};
    \draw[thick,dotted] (6,2.5) rectangle ++(1.92,1.4) node[rectnode,text width=2cm] {Clif. Tabl. $\ctabl_l$};
    \node at (8.4,3.2) {\Large$\cdots$};

    \node at (0.2,4.45) {\scriptsize\circled{2}};
    \path[fat arrow, thick] (1.5,4.1) to (2.1,4.8);
    \path[fat arrow, thick] (3.0,4.1) to (2.4,4.8);
    \node[align=center] at (2.25,4.3) {\scriptsize merge};

    \draw[thick] (0,5) rectangle ++(0.92,1.4) node[rectnode,text width=0.85cm] {CX+T Ckt $\mathcal{C}_{i1}$};
    \draw[thick,dotted,fill=gray!30] (1,5) rectangle ++(2.92,1.4) node[rectnode,text width=2cm] {Clif. Tabl. $\ctabl_j'$};
    \draw[thick,dashed] (4,5) rectangle ++(1.92,1.4) node[rectnode,text width=2cm] {PR. Tabl. $\ptabl_k$};
    \draw[thick,dotted] (6,5) rectangle ++(1.92,1.4) node[rectnode,text width=2cm] {Clif. Tabl. $\ctabl_l$};
    \node at (8.4,5.7) {\Large$\cdots$};

    \node at (0.2,6.95) {\scriptsize\circled{3}};
    \path[fat arrow, thick] (2.5,6.6) to (2.5,7.3);
    \node[align=center, text width=4.7cm] at (4,6.95) {\scriptsize T-count optimization +\\QC extraction/resyn};

    \draw[thick] (0,7.5) rectangle ++(0.92,1.4) node[rectnode,text width=0.85cm] {CX+T Ckt $\mathcal{C}_{i1}$};
    \draw[thick] (1,7.5) rectangle ++(0.92,1.4) node[rectnode,text width=0.85cm] {H-free Clif. Circuit $\mathcal{C}_{j1}$};
    \draw[thick] (2,7.5) rectangle ++(0.52,1.4) node[rectnode,text width=0.85cm] {Hs Ckt $\mathcal{C}_{j2}$};
    \draw[thick,dotted,fill=gray!30] (2.60,7.5) rectangle ++(1.32,1.4) node[rectnode,text width=1.16cm] {Clif. Tabl.\\$\ctabl_j''$\\{\tiny(H-free)}};
    \draw[thick,dashed] (4,7.5) rectangle ++(1.92,1.4) node[rectnode,text width=2cm] {PR. Tabl. $\ptabl_k$};
    \draw[thick,dotted] (6,7.5) rectangle ++(1.92,1.4) node[rectnode,text width=2cm] {Clif. Tabl. $\ctabl_l$};
    \node at (8.4,8.2) {\Large$\cdots$};

    \node at (0.2,9.45) {\scriptsize\circled{4}};
    \path[fat arrow,thick] (3.46,9.1) to (4.9,9.8);
    \path[fat arrow,thick] (4.8,9.1) to (3.36,9.8);
    \node[align=center] at (5.2,9.45) {\scriptsize swap};
    
    \draw[thick] (0,10) rectangle ++(0.92,1.4) node[rectnode,text width=0.85cm] {CX+T Ckt $\mathcal{C}_{i1}$};
    \draw[thick] (1,10) rectangle ++(0.92,1.4) node[rectnode,text width=0.85cm] {H-free Clif. Circuit $\mathcal{C}_{j1}$};
    \draw[thick] (2,10) rectangle ++(0.52,1.4) node[rectnode,text width=0.85cm] {Hs Ckt $\mathcal{C}_{j2}$};
    \draw[thick,dashed,fill=gray!30] (2.60,10) rectangle ++(1.92,1.4) node[rectnode,text width=2cm] {PR. Tabl. $\ptabl_k'$};
    \draw[thick,dotted] (4.60,10) rectangle ++(1.32,1.4) node[rectnode,text width=1.16cm] {Clif. Tabl.\\$\ctabl_j''$\\{\tiny(H-free)}};
    \draw[thick,dotted] (6,10) rectangle ++(1.92,1.4) node[rectnode,text width=2cm] {Clif. Tabl. $\ctabl_l$};
    \node at (8.4,10.7) {\Large$\cdots$};

    \node at (0.2,11.95) {\scriptsize\circled{5}};
    \path[fat arrow, thick] (3.9,11.6) to (3.9,12.3);
    \node[align=center, text width=4.7cm] at (5.4,11.95) {\scriptsize T-count optimization +\\QC extraction/resyn};

    \draw[thick] (0,12.5) rectangle ++(0.92,1.4) node[rectnode,text width=0.85cm] {CX+T Ckt $\mathcal{C}_{i1}$};
    \draw[thick] (1,12.5) rectangle ++(0.92,1.4) node[rectnode,text width=0.85cm] {H-free Clif. Circuit $\mathcal{C}_{j1}$};
    \draw[thick] (2,12.5) rectangle ++(0.52,1.4) node[rectnode,text width=0.85cm] {Hs Ckt $\mathcal{C}_{j2}$};
    \draw[thick] (2.60,12.5) rectangle ++(0.92,1.4) node[rectnode,text width=0.85cm] {CX+T Ckt $\mathcal{C}_{k1}$};
    \draw[thick,dotted,fill=gray!30] (3.60,12.5) rectangle ++(0.92,1.4) node[rectnode,text width=0.85cm] {Clif. Tabl. $\ctabl_{k2}$ {\tiny (CX-only)}};
    \draw[thick,dotted,fill=gray!30] (4.60,12.5) rectangle ++(1.32,1.4) node[rectnode,text width=1.16cm] {Clif. Tabl.\\$\ctabl_j''$\\{\tiny(H-free)}};
    \draw[thick,dotted,fill=gray!30] (6,12.5) rectangle ++(1.92,1.4) node[rectnode,text width=2cm] {Clif. Tabl. $\ctabl_l$};
    \node at (8.4,13.3) {\Large$\cdots$};

    \node at (0.2,14.45) {\scriptsize\circled{6}};
    \path[fat arrow, thick] (4.1,14.1) to (4.6,14.8);
    \path[fat arrow, thick] (5.22,14.1) to (5.22,14.8);
    \path[fat arrow, thick] (6.4,14.1) to (5.9,14.8);
    \node[align=center] at (5.7,14.3) {\scriptsize merge};
    
    \draw[thick] (0,15) rectangle ++(0.92,1.4) node[rectnode,text width=0.85cm] {CX+T Ckt $\mathcal{C}_{i1}$};
    \draw[thick] (1,15) rectangle ++(0.92,1.4) node[rectnode,text width=0.85cm] {H-free Clif. Circuit $\mathcal{C}_{j1}$};
    \draw[thick] (2,15) rectangle ++(0.52,1.4) node[rectnode,text width=0.85cm] {Hs Ckt $\mathcal{C}_{j2}$};
    \draw[thick] (2.60,15) rectangle ++(0.92,1.4) node[rectnode,text width=0.85cm] {CX+T Ckt $\mathcal{C}_{k1}$};
    \draw[thick,dotted,fill=gray!30] (3.60,15) rectangle ++(4.32,1.4) node[rectnode] {Clif. Tabl. $\ctabl_l'$};
    \node at (8.4,15.7) {\Large$\cdots$};

    \node at (0.2,16.95) {\scriptsize\circled{7}};
    \path[fat arrow, thick] (4.85,16.6) to (4.85,17.3);
    \node[align=center, text width=4.7cm] at (6.35,16.95) {\scriptsize T-count optimization +\\QC extraction/resyn};
    
    \draw[thick] (0,17.5) rectangle ++(0.92,1.4) node[rectnode,text width=0.85cm] {CX+T Ckt $\mathcal{C}_{i1}$};
    \draw[thick] (1,17.5) rectangle ++(0.92,1.4) node[rectnode,text width=0.85cm] {H-free Clif. Circuit $\mathcal{C}_{j1}$};
    \draw[thick] (2,17.5) rectangle ++(0.52,1.4) node[rectnode,text width=0.85cm] {Hs Ckt $\mathcal{C}_{j2}$};
    \draw[thick] (2.60,17.5) rectangle ++(0.92,1.4) node[rectnode,text width=0.85cm] {CX+T Ckt $\mathcal{C}_{k1}$};
    \draw[thick] (3.6,17.5) rectangle ++(0.92,1.4) node[rectnode,text width=0.85cm] {H-free Clif. Circuit $\mathcal{C}_{l1}$};
    \draw[thick] (4.6,17.5) rectangle ++(0.52,1.4) node[rectnode,text width=0.85cm] {Hs Ckt $\mathcal{C}_{l2}$};
    \draw[thick,dotted,fill=gray!30] (5.2,17.5) rectangle ++(1.32,1.4) node[rectnode] {Clif. Tabl. $\ctabl_l''$\\{\tiny (H-free)}};
    \node at (6.92,18.3) {\Large$\cdots$};

\end{tikzpicture}

%% file: src/4-experiments.tex
\section{Experiments} \label{sec:experiments}
\begin{table}[tb]
    \centering
    \caption{Algorithms and tools used for experiments}
    \begin{tabular}{l|c|l}
        \toprule
        Algorithm & Impl. & Configurations/Commands \\
        \midrule 
        Eager/Lazy & Qsyn & As detailed in \cref{sec:method} \\
        ZX & Qsyn & Full reduction \\
        & & $\to$ Extraction optimization level 2 \\
        Feynopt & Feynman & \texttt{-phasefold} \texttt{-cnotmin} \texttt{-simplify} \\
        \midrule \midrule
        moveH & Qsyn & - \\
        TKET & TKET & \texttt{RemoveRedundancies()} \\
        & & $\to$ \texttt{CliffordSimp()} \\
        & & $\to$ \texttt{RemoveRedundancies()} \\
        CausalFlowOpt & Qsyn & Max unfusions: 2; Max spider arity: 20\\
        \bottomrule
    \end{tabular}
    \label{tab:routine-detail}
\end{table}
\subsection{Experimental Settings}
We present the experimental results of our proposed algorithm in this section. All experiments were conducted on an Ubuntu 22.04 workstation equipped with a 5.5GHz Intel\textsuperscript{\textregistered} Core\textsuperscript{\texttrademark} i9-13900K CPU and 128GB of RAM. 

The proposed lazy resynthesis approach is implemented within Qsyn \cite{lau2024qsyn,qsynrepo}, our open-source quantum circuit synthesis framework that supports modern tableau-based, phase-polynomial-based, and ZX-calculus-based optimization techniques. For T-count optimization and Pauli rotation synthesis, we use the state-of-the-art FastTODD \cite{vandaele2024lower} and MST-based resynthesis \cite{vandaele2022phase}, respectively. To enable a fair comparison, we also implement the \textit{eager} resynthesis approach under the same QCO flow. The detailed configurations for all routines are summarized in \cref{tab:routine-detail}.

For benchmarks, we adopted the circuits used in \cite{vandaele2024lower}, a standard dataset for evaluating Clifford+T QCO algorithms. However, we excluded the \textit{hwb11} and \textit{hwb12} circuits from our benchmarks as FastTODD cannot terminate on these circuits within a time limit of 24 hours on our machine.

\subsection{Comparison on 2Q-Counts}

\begin{table}[tb]
    \centering
    \setlength\tabcolsep{2pt}
    \caption{Comparison on 2Q-count}
    \begin{tabular}{l|r|r|r|rrr|r}
    \toprule
    & \textbf{Preopt} & \multicolumn{2}{c|}{\textbf{Lazy}} & \textbf{Eager} & \textbf{ZX} & \textbf{Feynopt} & \textbf{FastTODD} \\\midrule
    \textbf{Benchmark} & \#2Q & \#2Q & \multicolumn{5}{c}{2Q-gate overhead\%} \\\midrule
\textbf{adder\_8} & 409 & 610 & 49.1 & 130.6 & \textbf{9.8} & 1072.6 & 942.1 \\
\textbf{barenco\_tof\_4} & 48 & 76 & 58.3 & 114.6 & \textbf{56.3} & 79.2 & 160.4 \\
\textbf{barenco\_tof\_5} & 72 & 131 & 81.9 & 150.0 & \textbf{48.6} & 105.6 & 186.1 \\
\textbf{barenco\_tof\_10} & 192 & 282 & 46.9 & 355.7 & \textbf{33.3} & 116.1 & 211.5 \\
\textbf{csla\_mux\_3} & 80 & 238 & 197.5 & 237.5 & \textbf{168.8} & 173.8 & 371.3 \\
\textbf{csum\_mux\_9} & 168 & 633 & 276.8 & 258.9 & \textbf{156.5} & 378.0 & 514.3 \\
\textbf{gf2\^{}6\_mult} & 221 & 477 & 115.8 & 114.9 & 319.9 & \textbf{102.3} & 479.6 \\
\textbf{gf2\^{}7\_mult} & 300 & 674 & \textbf{124.7} & 155.7 & 402.0 & 125.0 & 332.0 \\
\textbf{gf2\^{}8\_mult} & 405 & 1033 & \textbf{155.1} & 184.2 & 453.8 & 168.9 & 700.2 \\
\textbf{gf2\^{}9\_mult} & 494 & 1291 & \textbf{161.3} & 179.1 & 524.5 & 182.8 & 625.1 \\
\textbf{ham15-low} & 236 & 454 & 92.4 & 180.9 & \textbf{69.9} & 572.9 & 704.7 \\
\textbf{ham15-med} & 534 & 726 & 36.0 & 228.1 & \textbf{12.2} & 216.7 & 311.6 \\
\textbf{ham15-high} & 2149 & 3539 & 64.7 & 149.7 & \textbf{18.8} & 486.6 & 690.6 \\
\textbf{hwb6} & 116 & 152 & \textbf{31.0} & 129.3 & 37.9 & 200.9 & 387.1 \\
\textbf{hwb8} & 7129 & 10239 & 43.6 & 243.9 & \textbf{18.9} & 817.0 & 998.1 \\
\textbf{hwb10} & 35170 & 61817 & 75.8 & 335.5 & \textbf{43.7} & 1396.9 & 1550.3 \\
\textbf{mod\_mult\_55} & 48 & 117 & 143.8 & 170.8 & 172.9 & \textbf{85.4} & 227.1 \\
\textbf{mod\_red\_21} & 105 & 204 & 94.3 & 216.2 & \textbf{56.2} & 285.7 & 387.6 \\
\textbf{qcla\_adder\_10} & 233 & 586 & 151.5 & 430.9 & \textbf{110.3} & 137.8 & 286.7 \\
\textbf{qcla\_com\_7} & 186 & 324 & 74.2 & 108.1 & \textbf{46.8} & 52.7 & 162.4 \\
\textbf{qcla\_mod\_7} & 382 & 1498 & 292.1 & 431.2 & \textbf{78.8} & 459.4 & 501.6 \\
\textbf{rc\_adder\_6} & 93 & 140 & 50.5 & 283.9 & \textbf{49.5} & 61.3 & 147.3 \\
\textbf{tof\_5} & 42 & 71 & 69.0 & 235.7 & \textbf{40.5} & 228.6 & 242.9 \\
\textbf{tof\_10} & 102 & 155 & \textbf{52.0} & 614.7 & 70.6 & 443.1 & 365.7 \\
\textbf{vbe\_adder\_3} & 70 & 97 & \textbf{38.6} & 65.7 & 41.4 & 94.3 & 168.6 \\\midrule
\textbf{AVERAGE} &  &  & \textbf{103.1} & 228.2 & 121.7 & 321.7 & 466.2 \\
    \bottomrule
    \end{tabular}
    \label{tab:2q-counts}
\end{table}

\cref{tab:2q-counts} presents the experimental results on 2Q-counts.\footnote{We elided the T-counts in \cref{tab:2q-counts} since the resynthesis routines only differ in 2Q-count optimization. The T-counts of the optimized circuits are the same as in \cite{vandaele2024lower}.} The 2nd and 3rd columns are the 2Q-counts of the pre-optimized and post-optimized circuits by our lazy resynthesis approach. We compare the results with the eager approach (termed \textit{eager}), ZX calculus implementation in Qsyn (termed \textit{ZX}), and Feynopt (termed \textit{Feynopt}) \cite{amy2018towards,amy2014polynomial,amy2019formal,amy2017cnot}. To better characterize how much the 2Q-counts are increased, we calculate the percentage of overhead on the number of 2Q gates. For example, let the 2Q-counts for \textit{lazy} and \textit{Preopt} be 610 and 409, respectively. Then, the 2Q-gate overhead percentage would be (610 - 409) / 409 = 49.1\%.

We take the average for all the cases, and the results show that our \textit{lazy} approach, \textit{eager}, \textit{ZX}, and \textit{Feynopt} produce 103.1\%, 228.2\%, 121.7\%, and 321.7\% extra 2Q gates when compared to the \textit{preopt} circuits. While our result, 103.1\%, implies room for further improvement, we have demonstrated reductions in overheads of 54.8\%, 15.3\%, and 68.0\% over other techniques techniques, respectively.

Note that we obtain exceptional 2Q-count reductions on circuits with high numbers of internal Hadamard gates, such as \textit{ham15} and \textit{hwb} circuits. These results echo our derivation in Section III-C, where we show that the 2Q-count reduction asymptotically depends on the H-counts. On the other hand, for circuits without H-gates (eg. \textit{gf}-circuits), our algorithm slightly improves the 2Q-counts as compared to the \textit{eager} strategy.

It is also noteworthy that for a plain T-count optimization implementation, that is, no further 2Q-count optimization is exploited, the increase on 2Q-count can be as high as 466.2\% (i.e., the plain FastTODD algorithm).

\subsection{Runtime Comparison}

\begin{table}[tb]
    \centering
    \caption{Runtime Comparison}
    \begin{tabular}{l|r|rrr}
    \toprule
     & \textbf{Lazy} & \textbf{Eager} & \textbf{ZX} & \textbf{Feynopt} \\\midrule
    \textbf{Benchmark} & Runtime & \multicolumn{3}{c}{Runtime ratio\%} \\\midrule
\textbf{adder\_8} & 0.19 & \textbf{0.11} & 10.1 & 2.05 \\
\textbf{barenco\_tof\_4} & \textbf{0.01} & 3.00 & 11.0 & 2.00 \\
\textbf{barenco\_tof\_5} & \textbf{0.01} & 4.00 & 17.0 & 2.00 \\
\textbf{barenco\_tof\_10} & \textbf{0.10} & 8.80 & 6.4 & 0.30 \\
\textbf{csla\_mux\_3} & \textbf{0.03} & 1.67 & 10.7 & 1.00 \\
\textbf{csum\_mux\_9} & \textbf{0.09} & 1.44 & 10.0 & 0.56 \\
\textbf{gf2\^{}6\_mult} & \textbf{0.08} & 2.38 & 11.1 & 0.38 \\
\textbf{gf2\^{}7\_mult} & \textbf{0.09} & 3.33 & 10.8 & 0.33 \\
\textbf{gf2\^{}8\_mult} & 0.26 & 1.81 & 4.3 & \textbf{0.15} \\
\textbf{gf2\^{}9\_mult} & 0.42 & 1.31 & 2.5 & \textbf{0.14} \\
\textbf{ham15-low} & \textbf{0.08} & 2.50 & 11.5 & 0.25 \\
\textbf{ham15-med} & 0.31 & 1.52 & 3.3 & \textbf{0.06} \\
\textbf{ham15-high} & \textbf{0.12} & 0.50 & 54.3 & 0.83 \\
\textbf{hwb6} & \textbf{0.01} & 1.00 & 38.0 & 2.00 \\
\textbf{hwb8} & \textbf{0.21} & 0.81 & 402.3 & 20.71 \\
\textbf{hwb10} & 2.06 & \textbf{0.79} & 3042.7 & 111.08 \\
\textbf{mod\_mult\_55} & \textbf{0.01} & 3.00 & 37.0 & 3.00 \\
\textbf{mod\_red\_21} & \textbf{0.04} & 2.25 & 21.3 & 1.50 \\
\textbf{qcla\_adder\_10} & 0.27 & 1.26 & 2.6 & \textbf{0.26} \\
\textbf{qcla\_com\_7} & \textbf{0.12} & 1.42 & 6.3 & 0.25 \\
\textbf{qcla\_mod\_7} & \textbf{0.20} & 2.20 & 5.8 & 0.60 \\
\textbf{rc\_adder\_6} & \textbf{0.04} & 4.75 & 2.5 & 0.50 \\
\textbf{tof\_5} & \textbf{0.02} & 2.50 & 8.5 & 1.00 \\
\textbf{tof\_10} & \textbf{0.11} & 7.00 & 4.4 & 0.36 \\
\textbf{vbe\_adder\_3} & \textbf{0.02} & 2.00 & 8.0 & 1.50 \\\midrule
\textbf{GEOMEAN RATIO} & 1 & 1.81 & 13.1 & \textbf{0.84} \\
    \bottomrule
    \end{tabular}
    \label{tab:time}
\end{table}

\cref{tab:time} demonstrates the runtime comparison of the resynthesis routines. Note that the T-count optimization algorithm FastTODD is excluded from this comparison, as it is not part of the resynthesis process. We report the CPU time ratios of the other techniques versus our \textit{lazy} approach. The results show that our QCO algorithm is 1.81$\times$ and 13.1$\times$ faster than \textit{eager} and \textit{ZX}, respectively. On the other hand, \textit{Feynopt} runs slightly faster than our \textit{lazy} approach at 0.84$\times$ on average. 

According to the analysis in \cref{subsect:complexity}, although the runtime complexity of our \textit{lazy} approach is about the same as that of the eager approach, the experimental results suggest that the runtime spent on tableaux merging and swapping is much less than that on 2Q-gate synthesis. Therefore, our approach can achieve significant runtime improvements.

While our algorithm substantially improves tableau-based QCO, the best 2Q-counts are mostly distributed between our \textit{lazy} resynthesis and ZX-calculus-based approaches. ZX-calculus has a slightly higher 2Q-count overhead (121.7\%), but the performance varies from benchmark to benchmark. In the meantime, both \textit{ZX} and \textit{Feynopt} suffers from severe runtime and memory issues as the benchmark size increases (e.g., \textit{hwb8} and \textit{hwb10}), while our lazy method achieves better runtime reduction.

Interestingly, the ZX flow also performs the best on circuits with high H-counts and tends to be less effective when the H-count is low, suggesting that the ZX flow may be similar in some aspects to our algorithm. Indeed, the ZX resynthesis flow also works in a somewhat lazy manner, extracting trivial gates first before handling non-trivial ones. When stuck, it tries to rewrite the diagrams minimally so that the trivial gate extraction can continue.

\subsection{Further 2Q-Count Reduction by Local Optimization Techniques}

\begin{figure}
    \centering
    \includegraphics[width=0.9\linewidth]{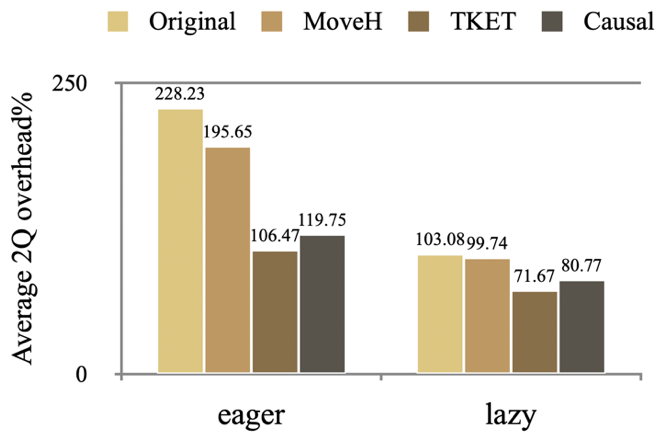}
    \caption{Further 2Q-count reduction by different local optimization techniques}
    \label{fig:local-opt}
\end{figure}

As shown in \ref{fig:comparison-t-2q}, the 2Q-count can be further improved by local optimization techniques \cite{de2020fast,holker2023causal,xu2022quartz}. We adopt the moveH algorithm in Qsyn, Remove Redundancies and Clifford Simplification from TKET, and the CausalFlowOpt techniques, as illustrated in \cref{tab:routine-detail}. We apply these methods to both \textit{lazy} and \textit{eager} approaches and present the 2Q-count improvements in \cref{fig:local-opt}.

%% file: src/5-conclusion.tex
\section{Conclusion} \label{sec:conclusion}

We have proposed a novel lazy resynthesis approach for modern tableau-based quantum circuit optimization (QCO) flows, substantially mitigating the 2Q-gate surges that commonly arise during T-count optimization in Clifford+T circuits. Experimental results demonstrate that our approach reduces 2Q-count overhead by 54.8\%, 15.3\%, and 68.0\% compared to tableau-based, ZX-calculus-based, and path-sum-based QCO algorithms, respectively. 
Additionally, our method achieves runtime speedups of 1.8$\times$ and 13.1$\times$ over the tableau-based and ZX-calculus-based methods, while performing comparably to the path-sum-based approach.

In conclusion, the proposed lazy resynthesis technique not only enhances the circuit quality and runtime performance of tableau-based QCO algorithms but also outper-forms alternative QCO frameworks such as ZX-calculus and path-sum-based methods in terms of both efficiency and scalability.

%% file: main.bbl
\begin{thebibliography}{10}
\providecommand{\url}[1]{#1}
\csname url@samestyle\endcsname
\providecommand{\newblock}{\relax}
\providecommand{\bibinfo}[2]{#2}
\providecommand{\BIBentrySTDinterwordspacing}{\spaceskip=0pt\relax}
\providecommand{\BIBentryALTinterwordstretchfactor}{4}
\providecommand{\BIBentryALTinterwordspacing}{\spaceskip=\fontdimen2\font plus
\BIBentryALTinterwordstretchfactor\fontdimen3\font minus \fontdimen4\font\relax}
\providecommand{\BIBforeignlanguage}[2]{{%
\expandafter\ifx\csname l@#1\endcsname\relax
\typeout{** WARNING: IEEEtran.bst: No hyphenation pattern has been}%
\typeout{** loaded for the language `#1'. Using the pattern for}%
\typeout{** the default language instead.}%
\else
\language=\csname l@#1\endcsname
\fi
#2}}
\providecommand{\BIBdecl}{\relax}
\BIBdecl

\bibitem{vandaele2024lower}
V.~Vandaele, ``Lower {$T$}-count with faster algorithms,'' \emph{arXiv preprint arXiv:2407.08695}, 2024.

\bibitem{ruiz2024quantum}
F.~J. Ruiz, T.~Laakkonen, J.~Bausch, M.~Balog, M.~Barekatain, F.~J. Heras, A.~Novikov, N.~Fitzpatrick, B.~Romera-Paredes, J.~van~de Wetering \emph{et~al.}, ``Quantum circuit optimization with alphatensor,'' \emph{arXiv preprint arXiv:2402.14396}, 2024.

\bibitem{heyfron2018efficient}
L.~E. Heyfron and E.~T. Campbell, ``An efficient quantum compiler that reduces {T} count,'' \emph{Quantum Science and Technology}, vol.~4, no.~1, p. 015004, 2018.

\bibitem{de2020fast}
N.~de~Beaudrap, X.~Bian, and Q.~Wang, ``Fast and effective techniques for {T}-count reduction via spider nest identities,'' \emph{arXiv preprint arXiv:2004.05164}, 2020.

\bibitem{bravyi2005universal}
S.~Bravyi and A.~Kitaev, ``Universal quantum computation with ideal {C}lifford gates and noisy ancillas,'' \emph{Physical Review A—Atomic, Molecular, and Optical Physics}, vol.~71, no.~2, p. 022316, 2005.

\bibitem{litinski2019magic}
D.~Litinski, ``Magic state distillation: Not as costly as you think,'' \emph{Quantum}, vol.~3, p. 205, 2019.

\bibitem{gidney2019efficient}
C.~Gidney and A.~G. Fowler, ``Efficient magic state factories with a catalyzed {$|CCZ\rangle$} to {$2|T\rangle$} transformation,'' \emph{Quantum}, vol.~3, p. 135, 2019.

\bibitem{campbell2017unifying}
E.~T. Campbell and M.~Howard, ``Unifying gate synthesis and magic state distillation,'' \emph{Physical review letters}, vol. 118, no.~6, p. 060501, 2017.

\bibitem{bravyi2012magic}
S.~Bravyi and J.~Haah, ``Magic-state distillation with low overhead,'' \emph{Physical Review A—Atomic, Molecular, and Optical Physics}, vol.~86, no.~5, p. 052329, 2012.

\bibitem{itogawa2024even}
T.~Itogawa, Y.~Takada, Y.~Hirano, and K.~Fujii, ``Even more efficient magic state distillation by zero-level distillation,'' \emph{arXiv preprint arXiv:2403.03991}, 2024.

\bibitem{gidney2024magic}
C.~Gidney, N.~Shutty, and C.~Jones, ``Magic state cultivation: growing {T} states as cheap as {CNOT} gates,'' \emph{arXiv preprint arXiv:2409.17595}, 2024.

\bibitem{zhang2019optimizing}
F.~Zhang and J.~Chen, ``Optimizing {$T$} gates in {C}lifford+{T} circuit as $\pi/4$ rotations around {P}aulis,'' \emph{arXiv preprint arXiv:1903.12456}, 2019.

\bibitem{kissinger2019reducing}
A.~Kissinger and J.~van~de Wetering, ``Reducing {T}-count with the {ZX}-calculus,'' \emph{arXiv preprint arXiv:1903.10477}, 2019.

\bibitem{vandaele2024optimal}
V.~Vandaele, S.~Martiel, S.~Perdrix, and C.~Vuillot, ``Optimal {H}adamard gate count for {C}lifford+{T} synthesis of {P}auli rotations sequences,'' \emph{ACM Transactions on Quantum Computing}, vol.~5, no.~1, pp. 1--29, 2024.

\bibitem{amy2014polynomial}
M.~Amy, D.~Maslov, and M.~Mosca, ``Polynomial-time {T}-depth optimization of {C}lifford+{T} circuits via matroid partitioning,'' \emph{IEEE Transactions on Computer-Aided Design of Integrated Circuits and Systems}, vol.~33, no.~10, pp. 1476--1489, 2014.

\bibitem{aaronson2004improved}
S.~Aaronson and D.~Gottesman, ``Improved simulation of stabilizer circuits,'' \emph{Physical Review A—Atomic, Molecular, and Optical Physics}, vol.~70, no.~5, p. 052328, 2004.

\bibitem{markov2008optimal}
K.~Markov, I.~Patel, and J.~Hayes, ``Optimal synthesis of linear reversible circuits,'' \emph{Quantum Information and Computation}, vol.~8, no. 3\&4, pp. 0282--0294, 2008.

\bibitem{maslov2018shorter}
D.~Maslov and M.~Roetteler, ``Shorter stabilizer circuits via bruhat decomposition and quantum circuit transformations,'' \emph{IEEE Transactions on Information Theory}, vol.~64, no.~7, pp. 4729--4738, 2018.

\bibitem{bravyi2021clifford}
S.~Bravyi, R.~Shaydulin, S.~Hu, and D.~Maslov, ``Clifford circuit optimization with templates and symbolic pauli gates,'' \emph{Quantum}, vol.~5, p. 580, 2021.

\bibitem{duncan2020graph}
R.~Duncan, A.~Kissinger, S.~Perdrix, and J.~Van De~Wetering, ``Graph-theoretic simplification of quantum circuits with the {ZX}-calculus,'' \emph{Quantum}, vol.~4, p. 279, 2020.

\bibitem{amy2017cnot}
M.~Amy, P.~Azimzadeh, and M.~Mosca, ``On the {CNOT}-complexity of {CNOT-PHASE} circuits,'' \emph{arXiv preprint arXiv:1712.01859}, 2017.

\bibitem{vandaele2022phase}
V.~Vandaele, S.~Martiel, and T.~G. de~Brugi{\`e}re, ``Phase polynomials synthesis algorithms for {NISQ} architectures and beyond,'' \emph{Quantum Science and Technology}, vol.~7, no.~4, p. 045027, 2022.

\bibitem{holker2023causal}
C.~Holker, ``Causal flow preserving optimisation of quantum circuits in the {ZX}-calculus,'' \emph{arXiv preprint arXiv:2312.02793}, 2023.

\bibitem{xu2022quartz}
M.~Xu, Z.~Li, O.~Padon, S.~Lin, J.~Pointing, A.~Hirth, H.~Ma, J.~Palsberg, A.~Aiken, U.~A. Acar \emph{et~al.}, ``Quartz: superoptimization of quantum circuits,'' in \emph{Proceedings of the 43rd ACM SIGPLAN International Conference on Programming Language Design and Implementation}, 2022, pp. 625--640.

\bibitem{lau2024qsyn}
M.-T. Lau, C.-Y. Cheng, C.-H. Lu, C.-H. Chuang, Y.-H. Kuo, H.-C. Yang, C.-T. Kuo, H.-Y. Chen, C.-Y. Tung, C.-E. Tsai \emph{et~al.}, ``{Q}syn: A developer-friendly quantum circuit synthesis framework for {NISQ} era and beyond,'' \emph{arXiv preprint arXiv:2405.07197}, 2024.

\bibitem{qsynrepo}
\BIBentryALTinterwordspacing
M.-T. Lau, C.-Y. Cheng, C.-H. Lu \emph{et~al.}, ``{Q}syn: A developer-friendly quantum circuit synthesis framework for {NISQ} era and beyond,'' [Open-source repository]. [Online]. Available: \url{https://github.com/DVLab-NTU/qsyn}
\BIBentrySTDinterwordspacing

\bibitem{cowtan2019phase}
A.~Cowtan, S.~Dilkes, R.~Duncan, W.~Simmons, and S.~Sivarajah, ``Phase gadget synthesis for shallow circuits,'' \emph{arXiv preprint arXiv:1906.01734}, 2019.

\bibitem{gottesman1998heisenberg}
D.~Gottesman, ``The {H}eisenberg representation of quantum computers,'' \emph{arXiv preprint quant-ph/9807006}, 1998.

\bibitem{meijer2024advances}
A.~Meijer-van~de Griend, ``Advances in quantum compilation in the nisq era,'' \emph{Doctoral Dissertation, University of Helsinki}, 2024.

\bibitem{amy2018towards}
M.~Amy, ``Towards large-scale functional verification of universal quantum circuits,'' \emph{arXiv preprint arXiv:1805.06908}, 2018.

\bibitem{amy2019formal}
------, ``Formal methods in quantum circuit design,'' Ph.D. dissertation, University of Waterloo Ontario, Canada, 2019.

\end{thebibliography}
